\def\bc{BRITE-Constellation}

\documentclass{iau}
\usepackage{graphicx}

\title[IAU Symp.301~~BRITE-Constellation] %% give here short title %%
{BRITE-Constellation: Nanosatellites for Precision Photometry of Bright Stars}

\author[Werner W. Weiss et al.] %% give here short author list %%
{W. W. Weiss$^{1\thanks{Member of the \bc\ Executive Science Team (BEST)}}$,
A. F. J. Moffat$^{2\dagger}$,
A. Schwarzenberg-Czerny$^{3\dagger}$,
\makebox{O. F. Koudelka$^{4\dagger}$},
C. C. Grant$^5$,
R. E. Zee$^5$,
R. Kuschnig$^{1\dagger}$,
\makebox{St. Mochnacki$^{6\dagger}$},
S. M. Rucinski$^{6\dagger}$,
J. M. Matthews$^{7\dagger}$,
P. Orleanski$^{8\dagger}$,
A. Pamyatnykh$^{3\dagger}$,
\makebox{A. Pigulski$^{9\dagger}$}, 
J. Alves$^{1\dagger}$,
\makebox{M. Guedel$^{1\dagger}$}, \\
G. Handler$^{3\dagger}$,
G. A. Wade$^{10\dagger}$,
A. L. Scholtz$^{11}$,
\makebox{\and the CCD Tiger Team\thanks{M. Chaumont, Cordell Grant, J. Lifshits, A. Popowicz, M. Rataj, P. Romano, M. Unterberger, R. Wawrzaszek, T. Zawistowski \& BEST}}
}

\affiliation{$^1$University of Vienna, Institute for Astrophysics, Tuerkenschanzstr. 17, 1180 Vienna, Austria; email: {\tt werner.weiss@univie.ac.at} \\ 
$^2$Dept. de physique, Universit$\acute e$ de Montr$\acute e$al, Canada;
$^3$Copernicus Astronomical Center, Warsaw, Poland;
$^4$Graz University of Technology. Graz, Austria;
$^5$Space Flight Laboratory, University of Toronto, Canada;
$^6$Dept. of Astronomy and Astrophysics, University of Toronto, Canada;
$^7$Dept. of Physics and Astronomy, University of British Columbia, Canada;
$^8$Space Research Center of the Polish Academy of Sciences, Warsaw, Poland;
$^9$Institute for Astronomy, University of Wroclaw. Poland; %Instytut Astronomiczny Uniwersytetu Wroclawskiego, Wroclaw, Poland
$^{10}$Dept. of Physics, Royal Military College of Canada, Ontario, Canada;
$^{11}$Institute of Telecommunications, Vienna University of Technology, Austria
}

\pubyear{2013}
\volume{301}  %% insert here IAU Symposium No.
\pagerange{119--126}
\jname{Precision Asteroseismology}
\editors{W. Chaplin, J. Guzik, G. Handler \& A. Pigulski eds.}
\begin{document}

\maketitle

\begin{abstract}
\bc\ (where BRITE stands for BRIght Target Explorer) is an international nanosatellite mission to monitor photometrically, in two colours, brightness and temperature variations of stars brighter than V$\approx 4$, with precision and time coverage not possible from the ground. The current mission design consists of three pairs of 7\,kg nanosats (hence "Constellation") from Austria, Canada and Poland carrying optical telescopes (3\,cm aperture) and CCDs.  One instrument in each pair is equipped with a blue filter; the other, a red filter.  The first two nanosats (funded by Austria), are UniBRITE, designed and built by UTIAS-SFL (University of Toronto Institute for Aerospace Studies – Spaceflight Laboratory) and its twin, BRITE-Austria, built by the Technical University Graz (TUG) with support of UTIAS-SFL. They were launched on 25 February 2013 by the Indian Space Agency, under contract to the Canadian Space Agency.
%, into a low-Earth dusk-dawn polar orbit.

Each BRITE instrument has a wide field of view ($\approx 24$ degrees), so up to 15 bright stars can be observed simultaneously in 32\,x\,32 sub-rasters.  Photometry (with reduced precision but thorough time sampling) of additional fainter targets will be possible through on-board data processing.  A critical technical element of the BRITE mission is the three-axis attitude control system to stabilize a nanosat with very low inertia.  The pointing stability is better than 1.5\,arcminutes rms, a significant advance by UTIAS-SFL over any previous nanosatellite.

\bc\ will primarily measure p- and g-mode pulsations to probe the interiors and ages of stars through asteroseismology. The BRITE sample of many of the brightest stars in the night sky is dominated by the most intrinsically luminous stars: massive stars seen at all evolutionary stages, and evolved medium-mass stars at the very end of their nuclear burning phases (cool giants and AGB stars).  The Hertzsprung-Russell Diagram  for stars brighter than magV=4 from which the \bc\ sample will be selected is shown in Fig.\,\ref{fig-hrd}. This sample falls into two principal classes of stars: 

(1) Hot luminous H-burning stars (O to F stars). Analyses of OB star variability have the potential to help solve two outstanding problems: the sizes of convective (mixed) cores in massive stars and the influence of rapid rotation on their structure and evolution. 

(2) Cool luminous stars (AGB stars, cool giants and cool supergiants). Measurements of the time scales involved in surface granulation and differential rotation will constrain turbulent convection models.

Mass loss from these stars (especially the massive supernova progenitors) is a major contributor to the evolution of the interstellar medium, so in a sense, this sample dominates cosmic ``ecology" in terms of future generations of star formation. The massive stars are believed to share many characteristics of the lower mass range of the first generation of stars ever formed (although the original examples are of course long gone). 

\begin{figure}[h] 
\center\includegraphics[width=0.75\textwidth]{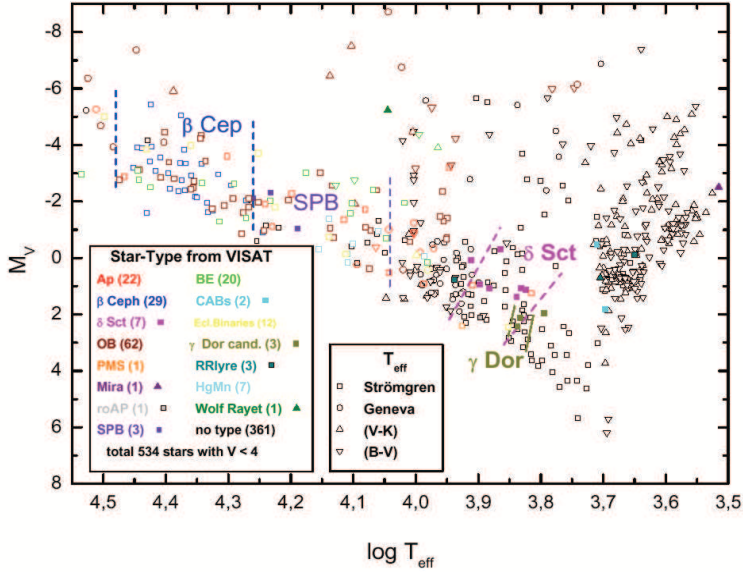}
%\vfill{5cm}
%\caption{Hertzsprung-Russell Diagram for stars brighter than mag(V)=4. Instability strips are indicated as well as the methods used to estimate the effective temperature. } 
\caption{HRD for stars brighter than magV\,=\,4, with instability strips indicated. } 
\label{fig-hrd}
\end{figure}

Other things that BRITE will do include detection of some Jupiter- and even Neptune-sized planets around bright host stars via transits, as expected on the basis of statistics from the Kepler exoplanet mission. Detecting planets around such very bright stars will greatly facilitate their subsequent characterization.  BRITE will also use surface spots to investigate stellar rotation.

%The next steps in completing \bc\ will be the launch of the first Polish nanosat, named “Lem” (after the Polish science fiction author Stanislaw Lem), which is scheduled for November 2013 aboard a Russian-Ukrainian DNEPR rocket. The second Polish nanosat, "Heweliusz", is slated for launch aboard a Chinese Long March LM-4 rocket. The two Canadian BRITE nanosats, called ``Toronto" and ``Montreal", are to be launched in 2014.

The following Table summarizes launch and orbit parameters of \bc\ components.
\vspace{-2mm}
\begin{table}[h]
\begin{tabular}{l|l|c|l|c|c|c|c}
%\hline
\hline
Designation & Name & F&Vehicle & T0 & Orbit & descending & Drift\\
                   &           &        &            &      &  km        & node          &  min \\
\hline
\multicolumn{2}{c|}{Owner: Austria} & & & & & & \\
BRITE-A & BRITE-Austria & B &PSLV-21 & 25Feb2013 & 800 & 18:00 & 0\\
BRITE-U & UniBRITE         & R &PSLV-21 & 25Feb2013 & circular & 18:00 & 0\\  
\multicolumn{2}{c|}{Owner: Canada} & & & & & & \\
BRITE-C1 & Toronto&       &DNEPR  & Q3-4/2014& 629x577& 10:30 & 40 \\
BRITE-C2 & Montr$\acute e$al&     & DNEPR  & Q3-4/2014& 629x577& 10:30 & 40 \\          
 \multicolumn{2}{c|}{Owner: Poland} & & & & & & \\
BRITE-P1  & Lem      & B    &DNEPR  &Nov.2013$^*$  & 600x900 & 10:30 & 100 \\
BRITE-P2  & Heveliusz& R   &China LM-4& Dec.2013$^*$ &    SSO/630\,km   &           &        \\
\hline
%\hline
\end{tabular}
%\caption{Entries for the Austrian payloads are based on the actual launch, and for the Canadian and Polish payloads are the planned parameters.  
%The "Designation" is the official satellite name according to the United Nations Register of objects launched into outer space; F...filter, T0...launch date ($^*$... as of August 2013); Orbit...height in km above ground; Drift...rate of drift from the initial ascending node in units of minutes per year.  The acronym "dd" stands for "dusk-dawn" orbit. }
\label{t:launch}
\end{table}
\vspace{-3mm}

The full version of this paper describing in more detail %the BRITE sample and science objectives, sky and time coverage, the two passbands used,  photometry at the BRITE focal plane,  BRITE hardware and mission operation, and commissioning of the first two nanosats, 
\bc\ will be published separately in a journal. The symposium presentation is available at \\ http://iaus301.astro.uni.wroc.pl/program.php 


\end{abstract}

%\begin{thebibliography}{}

%\end{thebibliography}

%\begin{discussion}

%\end{discussion}

\end{document}